\documentclass[acmlarge]{acmart}
\usepackage{multirow,tabularx}
\usepackage{soul}

\AtBeginDocument{%
  \providecommand\BibTeX{{%
    \normalfont B\kern-0.5em{\scshape i\kern-0.25em b}\kern-0.8em\TeX}}}

\setcopyright{acmcopyright}
\copyrightyear{2021}
\acmYear{2021}
\acmDOI{10.xxxx/xxxxx.xxxxx}

\acmJournal{JOCCH}
\begin{document}

\title{Providing More Efficient Access To Government Records: \\ A Use Case Involving Application of Machine Learning to Improve FOIA Review for the Deliberative Process Privilege}

\author{Jason R. Baron}
\affiliation{%
  \institution{University of Maryland}
  \city{College Park}
  \state{Maryland}
  \country{USA}}
\email{jrbaron@umd.edu}

\author{Mahmoud F. Sayed}
\affiliation{%
  \institution{University of Maryland}
  \city{College Park}
  \state{Maryland}
  \country{USA}}
\email{mfayoub@cs.umd.edu}

\author{Douglas W. Oard}
\affiliation{%
  \institution{University of Maryland}
  \city{College Park}
  \state{Maryland}
  \country{USA}}
\email{oard@umd.edu}

\renewcommand{\shortauthors}{Baron, et al.}
\renewcommand{\shorttitle}{Providing More Efficient Access To Gov. Records}

\begin{abstract}

At present, the review process for material that is exempt from disclosure under the Freedom of Information Act (FOIA) in the United States of America, and under many similar government transparency regimes worldwide, is entirely manual.  Public access to the records of their government is thus inhibited by the long backlogs of material awaiting such reviews.  This paper studies one aspect of that problem by first creating a new public test collection with annotations for one class of exempt material, the deliberative process privilege, and then by using that test collection to study the ability of current text classification techniques to identify those materials that are exempt from release under that privilege.  Results show that when the system is trained and evaluated using annotations from the same reviewer that even difficult cases can often be reliably detected, but that differences in reviewer interpretations, differences in record custodians, and that differences in topics of the records used for training and testing pose additional challenges.

\end{abstract}




\maketitle

\section{Introduction}
Early on in the COVID-19 pandemic, UNESCO recognized that the ``way the world is responding to this unprecedented global crisis will be part of history books.''  UNESCO went on to say that ``[m]emory institutions, including national archives \ldots are already recording the decisions and actions being made which will help future generations to understand the extent of the pandemic and its impact on societies,'' and that ``it is important, now more than ever, for memory institutions to become even more readily accessible to researchers, policymakers … and the community at large''~\cite{unesco2020covid}.

In the United States, the Freedom of Information Act (FOIA)\footnote{Title 5, U.S. Code, Section 552.}  was enacted in 1966 with the goal of fostering government accountability and transparency, by providing citizens with the opportunity to access the documentary heritage of the country.  Under the FOIA, all records created or received by federal agencies are presumptively open to public access, subject to nine enumerated statutory exemptions that under law are to be construed narrowly.\footnote{5 U.S.C. 552(b).}  In recent decades, FOIA has allowed the public to request records in electronic form, including e-mail and word processing documents.\footnote{See Electronic Freedom of Information Act Amendments of 1996, Pub. L. 104-231.}

However, as many critics have noted, a systemic weakness of FOIA administration is delays in response time to requestors~\cite{grunewald1998foia, weychert2016delayed}.  In large part, this is due to the near-uniform practice of FOIA officers engaging in manual searches for responsive records, coupled with even more time-consuming manual review of putatively responsive records to withhold (redact) material considered to be covered under a FOIA exemption.  A recent example of where delays in the FOIA process have had a measurable negative impact is in finding information on the government’s response to COVID-19. A number of federal agencies have been involved in the government’s response in managing COVID-19.\footnote{These include the Centers for Disease Control (CDC) and the National Institute of Health (NIH), as part of the White House Coronavirus Task Force; the State Department, in its dealings with the World Health Organization; and the Federal Emergency Management Agency’s efforts in providing emergency public assistance to state, local and tribal government entities.}  At the same time, subsequent to the initial outbreak of the pandemic in the United States in March 2020, the U.S. government faced extraordinarily severe criticism regarding their overall efforts to contain the spread of the virus.  During this moment in US history, federal agencies have seen a ``steep increase'' in the number of FOIA requests received for information on how the government has been responding to the crisis~\cite{foia2020covid}.  Given the fact that the federal government receives on the order of 800,000 FOIA requests on an annual basis~\cite{summary2019foia}, it is not surprising that agencies might be overwhelmed, causing substantial delays to requestors.  Indeed, during 2020 public interest groups sued agencies not only to receive specific records, but also for delaying the processing of FOIA requests due to COVID-19~\cite{thomsen2020watchdog}.

FOIA Exemption 5 allows the withholding of ``inter-agency or intra-agency memorandums or letters which would not be available by law to a party other than an agency in litigation with the agency.''   Exemption 5 allows agencies to justify withholding records through invocation of the ``deliberative process privilege.''\footnote{FOIA Exemption 5 also covers other forms of material exempt from disclosure, including records covered by the attorney-client privilege or constituting attorney work product.  Some exempt material may be exempt from disclosure for multiple reasons.  Here, we solely have limited our scope to evaluating textual material within the scope of the deliberative process privilege, irrespective of whether it may also be withholdable on other grounds}   The purpose of this privilege is ``to encourage honest and frank communication within the agency without fear of public disclosure.''\footnote{American Center for Law and Justice v. DOJ, 325 F.Supp.3d 162 (D.D.C. 2018).}   To satisfy being considered within the scope of the ``deliberative process privilege'' a record must in whole or in part meet two conditions: first, it must be ``predecisional,'' i.e., ``antecedent to the adoption of agency policy.''\footnote{Ancient Coin Collectors Guild v. U.S. Dep’t of State, 641 F.3d 504, 513 (D.C. Cir. 2011) (quoting prior case law).}  Second, the record  in whole or in part must be ``deliberative,'' i.e., ``when it reflects the give and take of decision making,'' including ``recommendations, draft documents, proposals, suggestions, and other subjective documents ``written by subordinate staff for consideration” by final decision makers.\footnote{Judicial Watch v. State Department, 349 F.Supp.3d 1, 7(D.D.C. 2018) (quoting in part Coastal States, 617 F.2d  854, 866 (D.C. Cir. 1980).} 

The FOIA statute also requires that ``any reasonably segregable portion of a record'' is releasable ``after deletion of the portions which are exempt.''\footnote{5 U.S.C.552(b).}   Under the statute, agencies shall ``consider whether partial disclosure of information is possible whenever the agency determines that a full disclosure of a requested record is not possible.''\footnote{5 U.S.C. 552(a)(8)(A)(ii).}  For purposes of the deliberative process privilege, a key consideration in determining whether material is segregable is whether the text of a document is discussing factual matters, as opposed to matters involving opinions, proposals, suggestions, and recommendations.

In acknowledgement of the increasingly archaic aspects of how FOIA searches are conducted throughout the government, a recent report written by an advisory council of FOIA experts recommended that the Archivist of the United States ``should work with other governmental components and industry in promoting research into using artificial intelligence, including machine learning technologies, to (i) improve the ability to search through government electronic recordkeeping repositories for responsive records to FOIA requests and (ii) identify sensitive material for potential segregation in government records, including but not limited to material otherwise within the scope of existing FOIA exemptions and exclusions''~\cite{ogis2020foia}. The experiments in this paper represent an initial attempt to meet the goal expressed in (ii) for one portion of one FOIA exemption.

This paper makes the following contributions:
\begin{itemize}
   
    \item We show that when classifiers are trained and tested under consistent conditions it is possible to design classifiers that achieve $F_1$ measures between 70\% and 83\% (i.e., if tuned so that precision and recall were equal, we would expect that between 70\% and 83\% of the exempt material would be found, and that the same fraction of the content identified by the classifier as exempt would truly be exempt).
    \item We study the effects of differences between reviewers, between the materials held by different custodians, and within the topical content of the records being classified to identify which differences pose the greatest challenge for current text classifiers.
    \item We control for the effect of document type and recognizable characteristics of content items to study classifier effectiveness on the content and document types that human reviewers find most difficult.
    \item We suggest directions for future work, identifying a need to model contextual factors that require access to evidence beyond the boundaries of specific documents.
    \item We introduce a new freely distributable test collection that is annotated for the deliberative process privilege under exemption 5 of the FOIA.\footnote{The test collection is available at \url{https://github.com/mfayoub/FOIA-Test-Collection}}
\end{itemize}

\section{Related Work}
There have been a numerous work in detecting sensitive content. One line of work is to redact named entities, e.g. person names, organizations, countries, etc. Additionally, it allows to redact special patterns, e.g. social security numbers, date, etc~\cite{abril2011declassification}. These techniques could be applied in different domains, e.g. medical records~\cite{sweeney1996replacing, tveit2004anonymization}. Document sanitization is one example that masks terms that are believed sensitive~\cite{sanchez2012detecting}. S{\'a}nchez et al. detects a sensitive term based on a measure called Information Content. This measure computes the probability of a term to appear in a corpus. A low probability of a term might indicate that the term is sensitive, and hence should be withheld from the containing document before release. In our work, one the classifiers predicts the sensitivity of a word based on the surrounding words. We think that our approach is more general and can adapt to different kinds of sensitivity.

So far as the authors are aware, to date there has not been research specifically aimed at applying machine learning techniques to segregate sensitive material in documents corresponding to any applicable U.S. FOIA exemptions.  However, within the U.K. there has been a substantial ongoing project through the SVGC Consortium to use machine learning techniques for assisting with sensitivity review and automated classification for information exempt under the U.K. Freedom of Information (FOI) Act (2000)~\cite{mcdonald2019framework}.  McDonald et al. used a test collection of 3,801 government documents, 502 (13.2\%) of which contained sensitive information pertaining to the categories of ``international relations,'' ``personal information,'' or both.  The authors studied the effect of lexical, syntactic and semantic features on classifier effectiveness, and conduced user studies to determine the utility of the resulting annotations to actual FOIA reviewers~\cite{mcdonald2015using, mcdonald2017enhancing}. Their test collection can not be redistributed because of the sensitive content that it contains. Unlike this work, we chose to work on paragraph-level as we think there are still non-sensitive paragraphs even inside sensitive documents that could be released. Also, we try different classifications algorithms.

There have also been two lines of work not specific to FOIA that nevertheless have yielded insights that could quite clearly be applied to various FOIA exemptions. The first is research regarding the detection of sensitive content in text~\cite{sayed2019jointly, sayed2020test}. Sayed et al. developed different search systems that try to find relevant, but not sensitive, content. These systems could be applied to protect personally identifiable information (PII), withholdable under FOIA Exemption 6~\cite{borden2016opening}.\footnote{Exemption 6 allows for withholding information in certain records which ``would constitute a clearly unwarranted invasion of personal privacy.''  5 U.S.C. 552(b)(6)} Complementing that work is research regarding detection of material covered under the attorney client privilege and attorney work product, for withholding in U.S. civil litigation~\cite{cormack2010overview,vinjumur2015finding,gabriel2013challenge,oard2018jointly}.  That research could be extended to detection of equivalent sensitivities under FOIA Exemption 5.

Finally, we note that although this paper focuses exclusively on detecting material that is exempt from release under the FOIA, there has been extensive work in the information retrieval field on the prerequisite task of finding material that requires review in response to a request~\cite{sayed2019jointly,grossman2016trec,kaczmarek2018email,cormack2015autonomy}.  There is also at least one test collection in which both relevant and sensitive content are annotated, although in that case the sensitivities are personal concerns rather than codified exemptions as in FOIA~\cite{sayed2020test}.   

%\doug{Mahmoud, we still need to flesh out this section.  Particularly Graham McDonald's work, but also more extensive references to related work generally.}

\section{Proposed Approach}

%\subsection{Modeling FOIA Review}

Our research approach was designed to model the workflow that agency staff follow in carrying out FOIA reviews to determine whether agency records should be disclosed or withheld from the public.
 
Upon receipt of a FOIA request, staff in a FOIA office decide which components of an agency are most likely to possess records responsive to the request. Sometimes the request needs to be clarified with the requestor to determine where in the agency to conduct a reasonable search.  A FOIA officer, working with both records managers and with individual custodians of records (e.g., employees), conducts a search that may include both hard-copy records in traditional files and electronic records.   In cases involving modest amounts of potentially responsive records, searches are carried out manually (and often inconsistently) by these individuals.  However, due to recent policy changes encouraging digital government, the volume of repositories of electronic records, and particularly e-mail and attachments, many agencies either are already or soon will be approaching in the millions of discrete records to be searched for FOIA requests~\cite{transition2019, archives2015capstone}.   

The authors are unaware of any federal agency that employs machine learning either for the purpose of conducting initial searches for responsive records in electronic repositories, or for finding exempt records or partially exempt records for purposes of determining what can be released pursuant to FOIA.   The initial search process is performed using some combination of manual searching and automated searching based on metadata (e.g., date or original custodians) and keywords, while a second round of sensitivity review for exempt material is conducted solely by manual means.  

With respect to finding textual material subject to the Exemption 5 deliberative process privilege, the manual review process consists of filtering using the following protocol:
\begin{enumerate}
    \item Is the record one where both the creator and all recipients are employed within the Executive branch?   Subject to minor exceptions, if the answer is “Yes” the record satisfies the Exemption 5 threshhold test for being “inter-agency” or “intra-agency” in nature.
    \item Are the entire contents of a particular record considered to be both “pre-decisional” and “deliberative”?
    \item If the answer to (2) is no, is there any portion (or portions) of the record that is considered to be “pre-decisional” and “deliberative”?   Ideally, determinations should be made based on a line-by-line review, but as a practical matter due to time and resources reviews often end up defaulting to decisions being made on a document-by-document basis at the first level of FOIA review, followed by a “paragraph by paragraph” basis on appeal.
\end{enumerate}

Initial FOIA review is conducted by FOIA office staff.  In cases where FOIA  determinations end up being subject to initial appeal, a different reviewer (often an attorney) will re-review records that have been withheld in whole or in part, for possible release. Requestors have the right to file a lawsuit if they continue to disagree with any non-disclosures of information.

Several aspects of the above protocol are of special importance to note.  First, the protocol described above consists of both easy and difficult tasks for a human reviewer.  Assume a newspaper article attached to an email has been identified in an initial search as potentially responsive to a given FOIA request due to a keyword appearing in the article.   A human reviewer would, however, immediately recognize the fact that a newspaper article is not an intra- or inter-agency document, and therefore would not be considered exempt.  For example, if the article were attached to an otherwise responsive and nonexempt email, both would be released to the requestor.   Optimally, a classifier must be trained to distinguish inter- and intra-agency records, from those arising and solely communicated within the Executive branch.  For our purposes here, we refer to categorically exempt documents as “Easy.”

Second, human reviewers have little or no difficulty in recognizing that some limited portions of agency records are “factual” in nature, outside the scope of what could possibly be deemed to be ``deliberative.'' This would be true for “Date,” “To, “From,” and “Subject” lines in a traditional document, as well as agency letterheads or other metadata.  For purposes of the experiments run here, we have chosen to designate this type of information in a document as ``trivial.''  For the same reason, we annotate header information and signature blocks in individual e-mail records as trivial since, with very limited exceptions, such information can not be withheld under exemption 5 (although it may be exempt from release under some other exemption).

Third, a human reviewer would have little difficulty segregating out “final” versions of records representing public releases of  agency positions or testimony or other statements that on the face are not “pre-decisional” or “deliberative” in nature.   Ideally, a classifier needs to be trained to recognize final versions of records, as opposed to drafts or other pre-decisional deliberations. For this research exercise, however, ``final'' versions of documents have been annotated as if they were drafts, since draft detection is not a research topic that we address here.

Finally, with respect to textual material of a substantive nature at the heart of the sensitivity review,  the review process places a priority on ensuring that all exempt material be properly identified.  This consideration significantly outweighs considerations of the volume of records to be reviewed. In other words, FOIA reviewers place the value of high recall over high precision, to ensure that textual material potentially subject to the deliberative process privilege is flagged. This also partially explains and contributes to delays in the overall FOIA process.  

%\jason{SEE DISCUSSION PARAGRAPHS WHICH INCORPORATE The NOTION THAT 0’s are more important than 1’s}

Aside from any difficulties posed for a classifier by the above considerations, the learning task involved here is made more difficult for two primary reasons.  First, this is due to interpretive nuances that have evolved over time in court decisions interpreting the scope of what constitutes ``predecisional'' and ``deliberative'' discussions,\footnote{See Judicial Watch v. Dep’t of State, 349 F.Supp.3d 1, 7 (D.D.C. 2018) (an ex post communication may still be predecisional if it discusses recommendations not expressly adopted).} as well as allowance for ad hoc determinations in restricted cases on what constitutes “reasonable segregability.”\footnote{For example, a court may decline to order an agency to commit significant time and resources to the separation of disjointed words, phrases, or even sentences which taken separately or together have minimal or no information content. See Mead Data Center v. Department of the Air Force, 566 F.2d 242, 261 n.55 (D.C. Cir. 1977).  In addition, when nonexempt information is “inextricably intertwined with exempt information, reasonable segregation is not possible.”  Id.} Second is that the degree of discretion that FOIA reviewers have in considering whether textual material falls in or outside of the scope of the exemption based on contextual considerations outside the four corners of a given document (e.g., see the discussion in Section~\ref{sec:annotation} of the “foreseeable harm” test).  Both considerations lead to differences in application amongst FOIA subject matter expert reviewers, and introduce some measure of difficulty in achieving ``gold standard'' certainty in training a classifier.  

%We therefore imposed certain artificial constraints on the classifier in its deciding whether records or portions thereof may be exempt from withholding under the deliberative process privilege.

\subsection{A Document Collection}
An initial difficulty encountered with this research was finding a suitable test collection.
Ideally, such a collection would be comprised of two version sets of documents  (1) an “original” set of documents previously withheld in whole or in part under Exemption 5, and (2) the same documents  now released or accessible to the public without the withheld portions. This was the approach taken by the Declassification Engine.\footnote{\url{https://web.archive.org/web/20130607100457/http://www.declassification-engine.org/redactions/\#/}} It proved difficult, however, to identify an adequately large collection of documents that either had been withheld under Exemption 5 but later released in litigation containing Exemption 5 withholdings, or a collection that had been withheld from public release due to its containing Exemption 5 material, but later reviewed and released after Exemption 5 concerns no longer applied due to the passage of time.\footnote{The deliberative process privilege ceases to be a basis for withholding after 25 years from creation date of an agency record.  5 U.S.C. 552(b)(5).}

An alternative strategy would have been to access selected archival records held at the National Archives and Records Administration (NARA), identifying through existing finding aids records once in senior agency officials’ files.  Such files would have a high probability of including policy option papers and other forms of consultation and recommendations, as part of the particular decisionmaking processes.  This would normally have been possible with some scanning and OCR, but the closure of public access to NARA's archival collections due to COVID-19 eliminated the possibility of pursuing this alternative.

A third option, however, presented itself in the form of searching through online record collections maintained on Presidential library websites.  Records covered by the Presidential Records Act (PRA) are accessible through FOIA five years after the end of a President’s term in office, subject to restrictions authorized by the statute that authorize a President to exempt records from disclosure for up to 12 years.  Under one of these restrictions (colloquially referred to as “P5”),  “confidential communications requesting or submitting advice, between the President and the President’s advisers, or between such advisers,” may be restricted for up to 12 years.\footnote{44 U.S.C. 2204(a)(5).}  Moreover, after the restriction period ends, such records cannot be withheld on the basis of Exemption 5.\footnote{Id., 2204(c).}  In other words, as a surrogate test collection, formerly restricted P5 records that have been opened in connection with FOIA requests would be excellent candidates for inclusion within this research, subject to understanding that there is not perfect congruence as a matter of law as between the P5 restriction and the deliberative process privilege.  As it turned out, we found many records where the P5 restriction had not been invoked, that nevertheless contained passages arguably withholdable under Exemption 5 had those records had been created in a federal agency subject to FOIA.  

%We elected to build the test collection from both formerly P5 restricted materials as well as from other records found in the overall Clinton presidential holdings. 

After further review, several presidential records collections accessible on the website of  the William J. Clinton Presidential Library appeared to meet the requisite requirements as outlined above.  The Clinton Library maintains its collections in designated “files,” each of which contains individual documents.\footnote{In keeping with the vocabulary of text classification, we refer to the items as documents.} A ``file'' consists of documents on a particular subject, similar to how a traditional fie cabinet would be organized by subject. Both files and documents are indexed by the White House component associated with their creation or receipt, as well as by the individual custodians who would have held the file.  In particular, a review of the records of now-Justice Elena Kagan, who in the 1990s worked as a lawyer on the Domestic Policy Council in the Clinton White House, contained numerous documents with formerly P5 restricted content that subsequently had been opened through FOIA requests or as part of the ongoing processing work of NARA archivists.  An additional search for records associated with Cynthia Rice, a second lawyer on the Domestic Policy Council,  were made part of the review. A keyword search for ``Elena Kagan'' produced 2,945 PDF files that contained varying numbers of documents.  From this we selected 32 files, a total of 432 documents, that on inspection were found to have a reasonable number of documents containing textual material arguably within the scope of the deliberative process privilege.  A keyword search for ``Cynthia Rice'' produced 1,072 files, from which we selected 5 files (a total of 77 documents) that met the same criterion. 

%\doug{Mahmoud: fill in XX above.}

\subsection{Annotation Protocols}
\label{sec:annotation}

Table~\ref{tab:collection} describes the collection of files used in the research.  “Batches” consist of an arbitrary number of subject matter files that were selected because they contained documents comprising a large number of pages.  The five batches of documents were reviewed sequentially in the order that they are numbered.  Four Batches (K1, K2, K3 and K5) were from files associated with Elena Kagan, consisting of a combined total of 32 files. An additional Batch R4 consisted of three Cynthia Rice files. For the first four batches, documents that the first reviewer felt were unlikely to contain material within the scope of the deliberative process privilege were skipped.  The entire fifth batch was annotated, however, including some documents that would easily be categorically excluded by a human reviewer. Documents from the fifth batch that the first reviewer felt could be categorically excluded from review were excluded from batch K5 and were instead used to used to form batch E5 (where E indicates an ``Easy'' decision). 

\begin{table}[h]
  \centering
    \begin{tabular}{|c|c|c|c|p{0.38\textwidth}|c|}
      \hline
      \textbf{Batch} & \textbf{Custodian} & \textbf{Files} & \textbf{Paragraphs} & \textbf{File Names} & \textbf{Reviewer(s)}\\
      \hline
      K1 & Elena Kagan & 9 & 523 & Superfund, Welfare Budget, Welfare-Blair Visit, Service Summit Policy, Service General, Veterans Affairs/Filipinos, Drugs Coerced Abstinence, Drugs Heroin Chic & A\\
      \hline
      K2 & Elena Kagan & 10 & 447 & Education/ TIMSS Meeting, Education/Troops to Teachers, Education/Vouchers, Environment/Climate Change, Kids Executive Order, Family Child Care Policy, Social Security/Nazis, Social Security/Prisoners, Drugs/Drug Testing & A \& B\\
      \hline
      K3 & Elena Kagan & 10 & 670 & Emails Received, Health/Radiation Experiments, Health/ Organ Transplants, Health/ Nursing Homes, Health/Medicaid Cap, Health/Immunization, Health/Genetic Screening, Drugs/Southwest Border, Environment/Port Dredging & A\\
      \hline
      R4 & Cynthia Rice & 5 & 466 & Child Support/Gambling, Child Support/License, Fathers/Bayh Bill, Budget 2001 FY New Ideas, Disability-Kennedy-Jeffords 1999 & A\\
      \hline
      K5 & Elena Kagan & 3 & 631 & Tax Proposals; Drugs/Media Campaign, Drugs/ Meth Report & A\\
      \hline
      E5 & Elena Kagan & 3 & 286 & Tax Proposals; Drugs/Media Campaign, Drugs/ Meth Report & A\\
      \hline
    \end{tabular}
    \caption{Document batches in the test collection.}
    \label{tab:collection}
  \centering
    \vspace{-6mm}%Put here to reduce too much white space after your table 
\end{table}

Individual documents in each batch were annotated by the lead author of this paper (Reviewer A), an attorney with subject matter expertise in FOIA law.  As detailed below, each document was divided into “paragraphs” for purposes of designation as either within or outside the scope of the exemption.  A second attorney with FOIA subject matter expertise (Reviewer B) also reviewed documents in Batch E2.  In the case of Batch K2, the two reviewers independently annotated documents in all 10 files, and then met together to review their annotations to see in cases of disagreement whether they could agree on a “consensus” annotation for training purposes.

Files in each batch ranged across a wide collection of subjects within the purview of the Domestic Policy Council, as indicated by the listed file names in Table~\ref{tab:collection}.  To support analysis of classifier performance on specific topics, Reviewer A assigned a single topic label to each document.  The list of topic labels assigned by Reviewer A is shown in Table~\ref{tab:file-topics-noneasy}.

A number of artificial constraints were placed on the attorney review process in order to simplify the core goal of finding pre-decisional and deliberative material within individual documents.  For the most part, these constraints reflect the fact some of the relevant criteria for making a real-world classification decision as to material within the scope of the deliberative process privilege are “contextual” in nature.   

First and foremost, as of 2016 the FOIA was amended to codify a “foreseeable harm” test, requiring agencies to withhold information “only if the agency reasonably foresees that disclosure would harm an interest protected by an exemption.”  Thus, passages that otherwise meet the test for what is considered “pre-decisional” and “deliberative” may nevertheless be required to be disclosed based on human judgment bringing a range of additional information about agency policies and current events to bear.   For purposes of this exercise, we did not annotate using a “foreseeable harm” test.  

Second, some initial scoping decisions were made so as not to muddle or make unduly difficult the classifier’s work.   Among these were: (a) parsing individual documents into paragraphs, for annotation purposes; (b) treating header information and signature block information (including in emails) as separate paragraphs from the main text; (c) annotating paragraphs as a whole as either exempt or not, rather than performing a more granular annotation on a sentence-by-sentence basis; (d) ignoring the presence or absence of ``DRAFT'' headers, instead treating all documents as ``drafts'' and thus per se within scope;  and (e) in Batches K1, K2, K3 and R4, excluding documents that Reviewer A found to be categorically nonexempt.   Examples of this nature included records originating outside the Executive branch, including from Congress or from outside lobbying groups; final legal briefs; final reports and white papers of various kinds; and newspaper and magazine articles.   We relaxed this last constraint (e) in reviewing the fifth Batch, where the reviewer annotated all documents found in three Kagan files.

Annotations were coded as follows: 

\begin{description}
\item[\ \ \ \ D1//] Paragraphs that fall within the scope of the deliberative process privilege
\item[\ \ \ \ E0//] ``Easy'' non-exempt paragraphs---those found in documents easily excluded from review due to not being "inter-agency" or "intra-agency" in nature (only annotated in batch E5)
\item[\ \ \ \ T0//]   ``Trivial'' non-exempt paragraphs (e.g., header information and signature blocks)
\item[\ \ \ \ D0//] ``Decided'' non-exempt paragraphs (i.e., paragraphs containing only factual content)
\end{description}

%\pagebreak

Figure~\ref{fig:example} shows an example of an annotated e-mail message.

\begin{figure}
\begin{verbatim}
  T0//
Sandra Thurman		01/12/98 10:35:44 AM Record Type:	Record
To:	Richard Socarides/WHO/EOP cc:	Maria Echaveste/WHO/EOP
Subject: Re: Needles/Embryos/Abortion and Other Selected L/HHS General Provisions SPEAK NOW OR...
  D1//
We did comment on the proposed language on needle exchange after consulting with both Chris
Jennings and Kevin Thurm. I will forward a copy of the memo to you.
  D1//
I had a lengthy discussion with Kevin last week regarding this issue. HHS does not plan to do
anything on needle exchange until Satcher is confirmed, assuming that will happen in February.
If indeed the confirmation is held up for some reason,we will have to revisit the
timing of any action.
  D1//
Contrary to what Scott Hitt may have told you, the AIDS community is still vehemently opposed 
to any law enforcement component in any compromise we might propose. So are General Mccaffrey
and I. In fact, it may well be the only point upon which we agree on this issue.
  0//
I am meeting again this week with the national AIDS groups to discuss where we are on needle
exchange. I'll keep you posted.
0//
Sandy
\end{verbatim}
\caption{An example of an annotated document.}
\label{fig:example}
\end{figure}

\subsection{Annotator Agreement}
We measured inconsistencies in annotation as between the two reviewers on batch K2, the only batch that was annotated by both reviewers (see Table~\ref{tab:agreement}). For this analysis we ignored T0 annotations (which were made only by Reviewer A).  The resulting Cohen's Kappa, 0.67, indicates substantial agreement.

\begin{table}[h]
  \centering
    \begin{tabular}{|l|c|c|c|c|}
      \hline
      & \multicolumn{4}{c|}{\textbf{Reviewer B}}\\
      \hline
      \multirow{4}{*}{\textbf{Reviewer A}} & & \textbf{D0} & \textbf{D1} & \textbf{Total}\\
      \cline{2-5}
      & \textbf{D0} & 212 & 69 & 281\\
      \cline{2-5}
      & \textbf{D1} & 6 & 160 & 166\\
      \cline{2-5}
      & \textbf{Total} & 218 & 229 & 447\\
      \hline
    \end{tabular}
    \caption{Annotator agreement, batch K2}
    \label{tab:agreement}
  \centering
    \vspace{-6mm}%Put here to reduce too much white space after your table 
\end{table}

During the consultation phase after initial annotations were completed, it became clear to the two attorney-reviewers that a substantial number of paragraphs where they disagreed in their assessments were found in documents known as “Talking Points,” or similar forms of “Qs and As,” related to an imminent public appearance of an official announcing government policy.  As it turns out, the inconsistent annotations reflect lack of settled precedent as to whether this type of document is ``predecisional'' or ``deliberative''.\footnote{Cf. American Center for Law and Justice v. US Dep’t of Justice, 325 F. Supp. 3d 162, 173 (D.D.C. 2018) (“A government employee drafting talking points . . . needs to know that her advice will remain privileged regardless of whether the [speaker] ultimately sticks to the script or decides to extemporize . . . Moreover, sticking to talking points often does not entail a verbatim recitation, leaving open the possibility that ‘a simple comparison’ of the talking points with the official’s public remarks would reveal the agency’s deliberations’), with Judicial Watch v. Dep’t of State, 349 F.Supp.3d 1, 7-8 (D.D.C. 2018) (``This discussion proves too much \ldots government officials give hundreds of speeches every day, all of which are important, though many elude recording or transcription.  So stretching the deliberative process privilege [to include talking points] would put many important public statements outside FOIA’s grasp, even well after the statements were made.'')}  Had  paragraphs in ``talking points'' memos been consistently decided, measured annotator agreement would have been higher.   

After computing annotator agreement, the two reviewers created an agreed final set of annotations for batch K2. We refer to these final annotations as having been created by ``Reviewer AB.''  We note that the reviewers reached complete consensus after the following revised determinations were made: Reviewer A, D0 to D1 (58), D1 to D0 (2); Reviewer B, D0 to D1 (4), D1 to D0 (11).  

%Among the 69 cases in which Reviewer A decided D0 and Reviewer B decided D1, 58 were resolved as D1 and 11 were resolved as D0. Among the 6 cases in which Reviewer A decided D1 and Reviewer B decided D0, 4 were resolved as D1 and 2 were resolved as D0.

\subsection{Evaluation Measure}

In the overall review task, performed by the human and machine working together, reviewers must be able to reach a high degree of confidence that all material that is exempt from release has been identified.  The machine's task is different, however, since the goal of the machine is to suggest, not to decide.  The machine's suggestions can be useful in two ways; they can help the human reviewer to avoid missing content exempt from release that they otherwise might miss, or they might help the human reviewer to decide more quickly on both exempt and non-exempt content.  To be useful, the machine must therefore find a substantial portion of the exempt content, and it must avoid misclassifying much of the non-exempt content as exempt.  With user studies we might be able to find the optimal balance between these two requirements, which correspond to recall and precision, respectively.  Our focus in this paper, however, is on characterizing the performance of existing classifiers for this task.  We therefore report recall and precision separately, and where a single objective is needed (as is the case when tuning parameters) we report the balanced harmonic mean of recall and precision ($F_1$). We compute 95\% confidence intervals for $F_1$ using normal approximation.  In result tables we bold the highest $F_1$ value and we underline values that are statistically significantly better than the ``All 1s'' baseline when the mean of each $F_1$ is outside the 95\% confidence interval of the other.

\subsection{Classifiers}

We seek to support human review for FOIA exemption 5 by highlighting paragraphs that the machine suggests may be exempt from release.  This is a binary classification in which each paragraph is to be given a label that indicates whether the paragraph is within the scope of the privilege (1 or positive class) or not (0 or negative class). Four approaches to text classification were employed for purposes of finding material subject to the deliberative process privilege.  These consisted of (1) Linear Regression (LR); (2) Support Vector Machine (SVM); (3) Begin-Inside-Outside (BIO) tagger using Conditional Random Fields; and (4) keyword search. 

In both LR and SVM, a raw paragraph is converted into a vector of word counts for the words in the paragraph. Then, this vector can be used as a training or test sample for classification. In BIO, each word in a document is annotated with either B, I, or O according to the following rules.
\begin{enumerate}
    \item If the current word is the beginning of privileged content, the word's label is B.
    \item If the current word is inside of privileged content, the word's label is I.
    \item If the current word is not part of privileged content, the word's label is O.
\end{enumerate}

The BIO classifier aims to predict the correct label of a word given its context, e.g. previous word and next word. Because our evaluation is based on paragraphs (as marked by Reviewer A), while the BIO classifier can place begin and end labels at any word, we need a way to map the BIO results to paragraph boundaries.  The BIO classifier therefore predicts a paragraph as privileged if the number of words predicted as B or I exceeds a certain percentage or non-privileged otherwise. This percentage is a hyper-parameter to the BIO classifier. 

As is obvious from the above, these factors can result in a huge number of combinations. As a result, we handle them in groups and present the results in the following subsections. It is worth noting that any training set is further split into training and validation sets. The validation set is used for hyper-parameters tuning. Table ... lists the hyper-parameters we tune and their corresponding ranges. We exhaustively consider all parameter combinations and extract the best parameters with the highest $F_1$ score on the validation set. Then we use the best parameters to estimate the class labels for test samples. 

\begin{table}[h]
  \centering
    \begin{tabular}{|l|l||l|l||l|l||}
      \hline
      \multicolumn{2}{|c||}{LR} & \multicolumn{2}{c||}{SVM} & \multicolumn{2}{c|}{BIO}\\
      \hline
      Parameter & Parameter Space & Parameter & Parameter Space & Parameter & Parameter Space\\
      \hline
      use\_idf & \{False, True\} & use\_idf & \{False, True\} & C1 & \{0.01, 0.1, 1, 5, 10\}\\
      stemmer & \{None, Porter\} & stemmer & \{None, Porter\} & C2 & \{0.01, 0.1, 1, 5, 10\}\\
      C & \{0.01, 0.1, 1, 5, 10\} & C & \{0.01, 0.1, 1, 5, 10\} & overlap \% & \{10, 20, ... 90, 100\}\\
      probability & \multirow{2}{*}{\{0.1, 0.2, ... 0.9, 1\}} & kernel & \{linear, rbf\}& & \\
      threshold & & $\gamma$ (for rbf) & \{1, 0.1, 0.01, 0.001, 0.0001\}  & & \\
      \hline
    \end{tabular}
    \caption{Parameter tuning by grid search.}
    \label{tab:hyper-parameters}
  \centering
    \vspace{-6mm}%Put here to reduce too much white space after your table 
\end{table}

Finally, keyword searches were performed using the following terms: \textit{option OR recommendation OR proposal OR suggest OR suggestion OR discuss OR discussion OR upcoming OR alternative OR frank OR candid OR ongoing}. If a paragraph contains any of those keywords, it was predicted to be privileged. If none of the keywords were found, the paragraph was predicted to be non-privileged.

In addition to the above four methods, we compared each against assignment of “all 1’s” (i.e., simply treating every paragraph as within the scope of the privilege). This simple approach achieves perfect recall, and is thus a useful baseline that other methods must exceed to be useful.

\section{Experiment Results}

We might imagine several scenarios in which text classification might be employed.  We have organized the presentation our our results around the four basic scenarios shown in Table~\ref{tab:custodian-attorney}.  These explore the following dimensions of variation:
\begin{itemize}
    \item Optimally the classifier would be trained and tested on annotations from the same reviewer, although that may not always be possible.
    
    \item Optimally the classifier would be trained and tested on documents that address a similar range of topics, although that may not always be possible.  In our experiments we use the custodian as a proxy for the set of topics on which that custodian worked. 
    
    \item We are most interested in the effectiveness of a classifier when it is presented with cases that call for the most difficult decisions, but it is also important that the classifier not make errors on the decisions that human reviewers find easy or trivial.

\end{itemize}

To explore these conditions, we conducted experiments for the conditions shown in Table \ref{tab:custodian-attorney}. As notation, we describe a condition by the reviewer followed by the batches.  for example, Train AB: K\textsubscript{2,3} would refer to training on Elena Kagan's batches 2 and 3, as annotated by the consensus of reviewers A and B.  We begin with experiments in which the task is to detect exempt paragraphs in documents that can not be categorically excluded (i.e., without the ``Easy'' documents in batch E5).

\begin{table}[h]
  \centering
    \begin{tabular}{|c|l|l|l|}
      \hline
      Condition & Documents & Reviewers & Method\\
      \hline
      A & Same custodian & Same reviewer & Cross-validation\\
      \hline
      B & Different custodians & Same reviewer & Train-test split\\
      \hline
      C & Same custodian & Different reviewers & Train-test split\\
      \hline
      D & Different custodians & Different reviewers & Train-test split\\
      \hline
    \end{tabular}
    \caption{Primary experiment conditions.}
    \label{tab:custodian-attorney}
  \centering
    \vspace{-6mm}%Put here to reduce too much white space after your table 
\end{table}

Table~\ref{tab:custodian-attorney-category-a} shows the results for Condition A, with the classifier trained on the same reviewer and the same custodian that it is tested on. All results are for 5-fold paragraph-scale cross-validation. Results, summarized in Table~\ref{tab:custodian-attorney-category-a}, show that the SVM performs well (by $F_1$).  There is no evidence in the results table favoring further consideration of the keyword classifier, at least with the specific keywords that we chose.

\begin{table}[h]
  \centering
    \begin{tabular}{|l|c|c|c||c|c|c||c|c|c||c|c|c|}
      \hline
       & \multicolumn{3}{l|}{\textbf{Train Cross-validate}} & \multicolumn{3}{l||}{\textbf{Train Cross-validate}} & \multicolumn{3}{l||}{\textbf{Train Cross-validate}} & \multicolumn{3}{l|}{\textbf{Train Cross-validate}} \\
       & \multicolumn{3}{l|}{\textbf{Test\;\,\, A: K\textsubscript{1,2,3,5}}} & \multicolumn{3}{l||}{\textbf{Test\;\,\, A: R\textsubscript{4}}} & \multicolumn{3}{l||}{\textbf{Test\;\,\, B: K\textsubscript{2}}} & \multicolumn{3}{l|}{\textbf{Test\;\,\, AB: K\textsubscript{2}}} \\
      \cline{2-13}
       & P & R & $F_1$ & P & R & $F_1$ & P & R & $F_1$ & P & R & $F_1$ \\
      \hline
      \textbf{LR} & 63.2 & 79.9 & \underline{\textbf{70.3$\pm$1.9}} & 62.5 & 76.4 & \underline{67.7$\pm$4.2} & 72.8 & 86.2 & \underline{78.2$\pm$3.8} & 77.5 & 83.9 & \underline{80.1$\pm$3.7} \\
      \hline
      \textbf{SVM} & 67.2 & 72.2 & \underline{69.5$\pm$1.9} & 70.4 & 72.2 & \underline{\textbf{70.7$\pm$4.1}} & 79.8 & 84.7 & \underline{\textbf{82.1$\pm$3.6}} & 80.2 & 87.2 & \underline{\textbf{83.5$\pm$3.4}} \\
      \hline
      \textbf{BIO} & 46.2 & 85.2 & \underline{59.8$\pm$2.0} & 46.8 & 94.2 & \underline{61.6$\pm$4.4} & 64.6 & 99.1 & \underline{78.2$\pm$3.8} & 64.9 & 98.3 & \underline{78.1$\pm$3.8} \\
      \hline
      \textbf{Keyword} & 53.0 & 32.1 & 39.9$\pm$2.0 & 56.9 & 38.8 & 45.0$\pm$4.5 & 86.2 & 29.3 & 43.1$\pm$4.6 & 86.2 & 29.8 & 43.5$\pm$4.6 \\
      \hline
      \textbf{All 1s} & 32.7 & 100 & 49.3$\pm$2.1  & 24.3 & 100 & 38.7$\pm$4.4 & 51.2 & 100 & 67.7$\pm$4.3 & 50.8 & 100 & 67.3$\pm$4.3 \\
      \hline
    \end{tabular}
    \caption{Condition A: same reviewer, same custodian, cross-validation; D1 vs \{D0, T0\}.}
    \label{tab:custodian-attorney-category-a}
  \centering
    \vspace{-6mm}%Put here to reduce too much white space after your table 
\end{table}

Table~\ref{tab:custodian-attorney-category-b} shows results for condition B, in which we trained classifiers using batches from one custodian annotated by one attorney, and then tested it using batches from a different custodian but annotated by the same attorney. As can be seen, the BIO classifier does well in this condition relative to other classifiers, but note that the $F_1$ values for the trained classifiers are well below the corresponding cross-validation values in Table~\ref{tab:custodian-attorney-category-a}.

\begin{table}[h]
  \centering
    \begin{tabular}{|l|c|c|c||c|c|c|}
      \hline
       & \multicolumn{3}{l||}{\textbf{Train A: K\textsubscript{1,2,3,5}}} & \multicolumn{3}{l|}{Train \textbf{Train A: R\textsubscript{4}}}\\
      
       & \multicolumn{3}{l||}{\textbf{Test\;\,\, A: R\textsubscript{4}}} & \multicolumn{3}{l|}{\textbf{Test\;\,\, A: K\textsubscript{1,2,3,5}}}\\
            \cline{2-7}
       & P & R & $F_1$ & P & R & $F_1$\\
      \hline
      \textbf{LR} & 47.0 & 61.9 & \underline{53.4$\pm$4.5} & 37.9 & 84.4 & \underline{52.3$\pm$2.1} \\
      \hline
      \textbf{SVM} & 43.7 & 83.2 & \underline{\textbf{57.3$\pm$4.5}} & 42.1 & 63.5 & 50.7$\pm$2.1 \\
      \hline
      \textbf{BIO} & 44.1 & 78.8 & \underline{56.5$\pm$4.5} & 39.6 & 83.7 & \underline{\textbf{53.8$\pm$2.1}} \\
      \hline
      \textbf{Keyword} & 56.8 & 37.2 & 44.9$\pm$4.5 & 52.9 & 32.0 & 39.9$\pm$2.0 \\
      \hline
      \textbf{All 1s} & 24.2 & 100 & 39.0$\pm$4.4 & 32.7 & 100 & 49.3$\pm$2.1 \\
      \hline
    \end{tabular}
    \caption{Condition B: same reviewer, different custodian; D1 vs \{D0, T0\}.}
    \label{tab:custodian-attorney-category-b}
  \centering
    \vspace{-6mm}%Put here to reduce too much white space after your table 
\end{table}

Table~\ref{tab:custodian-attorney-category-c} shows results for condition C, in which we have trained our classifiers using batches from one custodian annotated by one reviewer, and then we test using batches from the same custodian, but annotated by a different reviewer. As was the case for condition A, the SVM classifier performs well (by $F_1$).

\begin{table}[h]
  \centering
    \begin{tabular}{|l|c|c|c||c|c|c||c|c|c||c|c|c|}
      \hline
       & \multicolumn{3}{l||}{\textbf{Train A: K\textsubscript{1,3,5}}} & \multicolumn{3}{l||}{\textbf{Train A: K\textsubscript{1,3,5}}} & \multicolumn{3}{l||}{\textbf{Train B: K\textsubscript{2}}} & \multicolumn{3}{l|}{\textbf{Train AB: K\textsubscript{2}}} \\
       & \multicolumn{3}{l||}{\textbf{Test\;\,\, B: K\textsubscript{2}}} & \multicolumn{3}{l||}{\textbf{Test\;\,\, AB: K\textsubscript{2}}} & \multicolumn{3}{l||}{\textbf{Test\;\,\, A: K\textsubscript{1,3,5}}} & \multicolumn{3}{l|}{\textbf{Test\;\,\, A: K\textsubscript{1,3,5}}} \\
      \cline{2-13}
       & P & R & $F_1$ & P & R & $F_1$ & P & R & $F_1$ & P & R & $F_1$ \\
      \hline
      \textbf{LR} & 78.3 & 31.4 & 44.9$\pm$4.6 & 76.1 & 30.8 & 43.9$\pm$4.6 & 44.8 & 63.8 & \underline{\textbf{52.6$\pm$2.3}} & 42.7 & 63.4 & \underline{51.0$\pm$2.3} \\
      \hline
      \textbf{SVM} & 66.8 & 93.9 & \underline{\textbf{78.0$\pm$3.8}} & 66.5 & 94.3 & \underline{\textbf{78.0$\pm$3.8}} & 36.5 & 91.0 & \underline{52.1$\pm$2.3} & 40.8 & 76.3 & \underline{\textbf{53.1$\pm$2.3}} \\
      \hline
      \textbf{BIO} & 66.8 & 86.9 & \underline{75.5$\pm$4.0} & 67.1 & 88.1 & \underline{76.2$\pm$3.9} & 39.1 & 80.2 & \underline{\textbf{52.6$\pm$2.3}} & 39.4 & 76.6 & \underline{52.0$\pm$2.3} \\
      \hline
      \textbf{Keyword} & 86.1 & 29.7 & 44.2$\pm$4.6 & 86.1 & 30.0 & 44.4$\pm$4.6 & 48.5 & 31.2 & 38.0$\pm$2.2 & 48.5 & 31.2 & 38.0$\pm$2.2 \\
      \hline
      \textbf{All 1s} & 51.2 & 100 & 67.8$\pm$4.3 & 50.8 & 100 & 67.4$\pm$4.3 & 31.7 & 100 & 48.1$\pm$2.3 & 31.7 & 100 & 48.1$\pm$2.3 \\
      \hline
    \end{tabular}
    \caption{Condition C: different reviewer, same custodian; D1 vs \{D0, T0\}.}
    \label{tab:custodian-attorney-category-c}
  \centering
    \vspace{-6mm}%Put here to reduce too much white space after your table 
\end{table}

Table~\ref{tab:custodian-attorney-category-d} shows results for category D, in which we trained using batches from one custodian annotated by some reviewer, and then tested using batches from a different custodian that were annotated by a different reviewer. As the results show, the Logistic Regression classifier consistently does well in this condition (as measured by $F_1$), with the BIO classifier not far behind.

\begin{table}[h]
  \centering
    \begin{tabular}{|l|c|c|c||c|c|c||c|c|c||c|c|c|}
      \hline
       & \multicolumn{3}{l||}{\textbf{Train A: R\textsubscript{4}}} & \multicolumn{3}{l||}{\textbf{Train A: R\textsubscript{4}}} & \multicolumn{3}{l||}{\textbf{Train B: K\textsubscript{2}}} & \multicolumn{3}{l|}{\textbf{Train AB: K\textsubscript{2}}} \\
       & \multicolumn{3}{l||}{\textbf{Test\;\,\, B: K\textsubscript{2}}} & \multicolumn{3}{l||}{\textbf{Test\;\,\, AB: K\textsubscript{2}}} & \multicolumn{3}{l||}{\textbf{Test\;\,\, A: R\textsubscript{4}}} & \multicolumn{3}{l|}{\textbf{Test\;\,\, A: R\textsubscript{4}}} \\
      \cline{2-13}
       & P & R & $F_1$ & P & R & $F_1$ & P & R & $F_1$ & P & R & $F_1$ \\
      \hline
      \textbf{LR} & 63.5 & 87.3 & \underline{\textbf{73.5$\pm$4.1}} & 63.5 & 88.1 & \underline{\textbf{73.8$\pm$4.1}} & 46.3 & 77.0 & \underline{\textbf{57.8$\pm$4.5}} & 46.2 & 79.6 & \underline{\textbf{58.4$\pm$4.5}} \\
      \hline
      \textbf{SVM} & 65.7 & 62.0 & 63.8$\pm$4.5 & 67.6 & 64.3 & 65.9$\pm$4.4 & 36.2 & 98.2 & \underline{52.9$\pm$4.5} & 42.1 & 90.3 & \underline{57.5$\pm$4.5} \\
      \hline
      \textbf{BIO} & 63.9 & 82.1 & \underline{71.9$\pm$4.2} & 63.9 & 82.8 & \underline{72.2$\pm$4.2} & 41.6 & 87.6 & \underline{56.4$\pm$4.5} & 42.4 & 84.1 & \underline{56.4$\pm$4.5} \\
      \hline
      \textbf{Keyword} & 86.1 & 29.7 & 44.2$\pm$4.6 & 86.1 & 30.0 & 44.4$\pm$4.6 & 56.8 & 37.2 & 44.9$\pm$4.5 & 56.8 & 37.2 & 44.9$\pm$4.5 \\
      \hline
      \textbf{All 1s} & 51.2 & 100 & 67.8$\pm$4.3 & 50.8 & 100 & 67.4$\pm$4.3 & 24.2 & 100.0 & 39.0$\pm$4.4 & 24.2 & 100.0 & 39.0$\pm$4.4 \\
      \hline
    \end{tabular}
    \caption{Condition D: different reviewer, different custodian, D1 vs \{D0, T0\}.}
    \label{tab:custodian-attorney-category-d}
  \centering
    \vspace{-6mm}%Put here to reduce too much white space after your table 
\end{table}

\subsection{Including Paragraphs from Easily Recognized Non-Exempt Documents}

Table~\ref{tab:easy-zeros} shows the effect of adding batch E5, which consists entirely of the documents in Elena Kagan's Batch 5 that a human reviewer would easily recognize as categorically non-exempt.  This has the effect of adding 286 additional paragraphs, all with E0 annotations, to the training set (in condition B), to the test set (in condition C), or to both (in condition A).  Adding batch E5 to the test set (as happens for Conditions A and C) reduces $F_1$ values, suggesting that documents that human reviewers can easily recognize as categorically non-exempt are not actually easy for classifiers that see only isolated paragraphs. This suggests that it may be useful to automatically recognize categorically non-exempt documents as a preprocessing stage, something we did not try in our experiments.  Interestingly, adding batch E5 to training (as happens for condition B) seems to have only small effects on $F_1$.

\begin{table}[h]
  \centering
    \begin{tabular}{|l|c|c|c||c|c|c||c|c|c|}
      \hline
       & \multicolumn{3}{l||}{\textbf{Train Cross-validate}} & \multicolumn{3}{l||}{\textbf{Train A: K\textsubscript{1,2,3,5}+E\textsubscript{5}}} & \multicolumn{3}{l|}{\textbf{Train B: K\textsubscript{2}}} \\
       & \multicolumn{3}{l||}{\textbf{Test\;\,\, A: K\textsubscript{1,2,3,5}+E\textsubscript{5}}} & \multicolumn{3}{l||}{\textbf{Test\;\,\, A: R\textsubscript{4}}} & \multicolumn{3}{l||}{\textbf{Test\;\,\, A: K\textsubscript{1,3,5}+E\textsubscript{5}}} \\
      \cline{2-10}
       & P & R & $F_1$ & P & R & $F_1$ & P & R & $F_1$ \\
      \hline
      \textbf{LR} & 61.7 & 74.7 & \underline{\textbf{67.4$\pm$1.8}} & 52.9 & 63.7 & \underline{57.8$\pm$4.5} & 40.0 & 63.8 & \underline{\textbf{49.2$\pm$2.1}} \\
      \hline
      \textbf{SVM} & 62.8 & 71.8 & \underline{66.9$\pm$1.8} & 45.6 & 77.9 & \underline{57.5$\pm$4.5} & 31.7 & 91.0 & \underline{47.0$\pm$2.1} \\
      \hline
      \textbf{BIO} & 45.8 & 71.4 & \underline{55.6$\pm$1.9} & 47.5 & 76.1 & \underline{\textbf{58.5$\pm$4.5}} & 34.0 & 80.2 & \underline{47.8$\pm$2.1} \\
      \hline
      \textbf{Keyword} & 51.0 & 32.1 & 39.3$\pm$1.9 & 56.8 & 37.2 & 44.9$\pm$4.5 & 46.6 & 31.2 & 37.4$\pm$2.1 \\
      \hline
      \textbf{All 1s} & 29.1 & 100 & 45.0$\pm$1.9 & 24.2 & 100 & 39.0$\pm$4.4 & 27.3 & 100 & 42.9$\pm$2.1 \\
      \hline
    \end{tabular}
    \caption{Effects of adding Batch E5; Left to right: one example each for conditions A, B, C; D1 vs \{D0, T0, E0\}}.
    \label{tab:easy-zeros}
  \centering
    \vspace{-6mm}%Put here to reduce too much white space after your table 
\end{table}

\subsection{Excluding Trivially Non-Exempt Paragraphs}

Table~\ref{tab:non-trivial} shows one representative result for each condition when trivially non-exempt paragraphs such as headers and signature blocks (i.e., those with T0 annotations) are excluded from both training and testing.  For condition A, in which cross-validation yields training and test conditions that are remarkable similar, the three trained classifiers yield results very similar to those of the corresponding condition in Table~\ref{tab:custodian-attorney-category-a}.  However, the best trained classifier results for conditions B, C and D, do not exceed ``all 1s'' baseline.  This occurs in part because the ``all 1s'' baseline naturally improves when any zero annotations (including T0) are excluded, and we expect in part because the resulting classification task is harder.

\begin{table}[h]
  \centering
    \begin{tabular}{|l|c|c|c||c|c|c||c|c|c||c|c|c|}
      \hline
      \multirow{3}{*}{\textbf{}} & \multicolumn{3}{l||}{\textbf{Train cross-validate}} & \multicolumn{3}{l||}{\textbf{Train A: R\textsubscript{4}}} & \multicolumn{3}{l||}{\textbf{Train A: K\textsubscript{1,3,5}}} & \multicolumn{3}{l|}{\textbf{Train A: R\textsubscript{4}}} \\
       & \multicolumn{3}{l||}{\textbf{Test\;\,\, A: K\textsubscript{1,2,3,5}}} & \multicolumn{3}{l||}{\textbf{Test\;\,\, A: K\textsubscript{1,2,3,5}}} & \multicolumn{3}{l||}{\textbf{Test\;\,\, AB: K\textsubscript{2}}} & \multicolumn{3}{l|}{\textbf{Test\;\,\, AB: K\textsubscript{2}}} \\
      \cline{2-13}
       & P & R & $F_1$$\pm$CI & P & R & $F_1$$\pm$CI & P & R & $F_1$$\pm$CI & P & R & $F_1$$\pm$CI \\
      \hline
      \textbf{LR} & 65.3 & 75.9  & \underline{70.1$\pm$2.0} & 38.9 & 48.9 & 43.3$\pm$2.2 & 79.9 & 57.0 & 66.5$\pm$4.9 & 67.2 & 72.2 & 69.6$\pm$4.8 \\
      \hline
      \textbf{SVM} & 72.0 & 68.9 & \underline{\textbf{70.4$\pm$2.0}} & 40.0 & 51.3 & 44.9$\pm$2.2 & 79.4 & 43.5 & 56.2$\pm$5.2 & 66.6 & 95.2 & 78.4$\pm$4.3 \\
      \hline
      \textbf{BIO} & 53.0 & 77.3 & \underline{62.2$\pm$2.1} & 41.0 & 73.0 & 52.5$\pm$2.2 & 75.9 & 70.0 & 72.9$\pm$4.7 & 67.6 & 55.2 & 60.8$\pm$5.1 \\
      \hline
      \textbf{Keyword} & 55.0 & 32.1 & 40.5$\pm$2.2 & 55.2 & 32.2 & 40.7$\pm$2.2 & 93.2 & 29.6 & 44.9$\pm$5.2 & 93.2 & 29.6 & 44.9$\pm$5.2 \\
      \hline
      \textbf{All 1s} & 37.8 & 100 & 54.8$\pm$2.2 & 37.8 & 100 & \textbf{54.9$\pm$2.2} & 65.7 & 100 & \textbf{79.3$\pm$4.2} & 65.7 & 100 & \textbf{79.3$\pm$4.2} \\
      \hline
    \end{tabular}
    \caption{Effects of exclusion of T0 (trivially non-exempt) paragraphs; left to right: conditions A, B, C, D; D1 vs D0.}
    \label{tab:non-trivial}
  \centering
    \vspace{-6mm}%Put here to reduce too much white space after your table 
\end{table}

\subsection{Topic Effects}

Table~\ref{tab:file-topics-noneasy} shows $F_1$ values for classifiers that were tested using Reviewer A's judgments for all paragraphs that had been labeled with the topic shown.  Each classifier was trained using Reviewer A's judgments for all other paragraphs (i.e., paragraphs labeled with any other topic).  As in Tables~\ref{tab:custodian-attorney-category-a} through~\ref{tab:custodian-attorney-category-d}, trivial zeros were included and easy zeros (i.e., batch E5) were excluded.  For 13 of the 16 topics, covering 95\% of all paragraphs (2,590 of 2,735), some trained classifier statistically significantly outperformed the ``all 1s'' baseline.  The keyword classifier exhibited the greatest variation, with $F_1$ values between 0.059 (for Education) and 0.600 (for Tax Proposals and for Family).

\begin{table}[h]
  \centering
    \begin{tabular}{|l|c|c c c c c|}
        \hline
        \textbf{Topic} & \textbf{Paragraphs} & \textbf{LR} & \textbf{SVM} & \textbf{BIO} & \textbf{Keyword} & \textbf{All 1s} \\
        \hline
        Drugs & 699 & 25.4$\pm$3.2 & 44.4$\pm$3.7 & \underline{\textbf{55.2$\pm$3.7}} & 22.4$\pm$3.1 & 51.6$\pm$3.7 \\
        \hline
        Health & 297 & \underline{\textbf{55.2$\pm$5.7}} & \underline{47.7$\pm$5.7} & 41.1$\pm$5.6 & 41.8$\pm$5.6 & 37.7$\pm$5.5 \\
        \hline
        Tax Proposals & 253 & 54.3$\pm$6.1 & \underline{\textbf{61.8$\pm$6.0}} & 57.7$\pm$6.1 & 60.0$\pm$6.0 & 54.6$\pm$6.1 \\
        \hline
        Welfare & 245 & \underline{\textbf{51.3$\pm$6.3}} & 39.3$\pm$6.1 & 38.5$\pm$6.1 & 31.5$\pm$5.8 & 38.3$\pm$6.1\\
        \hline
        Child Support & 216 & \underline{\textbf{59.3$\pm$6.6}} & 44.0$\pm$6.6 & 39.3$\pm$6.5 & 54.0$\pm$6.6 & 29.2$\pm$6.1 \\
        \hline
        Service & 198 & \underline{\textbf{62.0$\pm$6.8}} & 56.6$\pm$6.9 & 54.0$\pm$6.9 & 40.0$\pm$6.8 & 53.9$\pm$6.9 \\
        \hline
        Miscellaneous Emails & 188 & \underline{\textbf{62.9$\pm$6.9}} & 50.7$\pm$7.1 & 43.8$\pm$7.1 & 44.8$\pm$7.1 & 33.6$\pm$6.8 \\
        \hline
        Disability & 105 & \underline{\textbf{61.0$\pm$9.3}} & 51.8$\pm$9.6 & 51.7$\pm$9.6 & 50.0$\pm$9.6 & 35.9$\pm$9.2 \\
        \hline
        Education & 103 & 37.9$\pm$9.4 & \underline{\textbf{53.2$\pm$9.6}} & 48.5$\pm$9.7 & 5.9$\pm$4.5 & 41.5$\pm$9.5 \\
        \hline
        Budget & 100 & 57.1$\pm$9.7 & \underline{\textbf{80.7$\pm$7.7}} & 70.5$\pm$8.9 & 40.0$\pm$9.6 & 63.0$\pm$9.5 \\
        \hline
        Kids & 92 & 69.2$\pm$9.4 & \underline{\textbf{91.9$\pm$5.6}} & 77.5$\pm$8.5 & 56.3$\pm$10.1 & 82.8$\pm$7.7 \\
        \hline
        Environment & 73 & 62.0$\pm$11.1 & \textbf{66.7$\pm$10.8} & 62.5$\pm$11.1 & 54.5$\pm$11.4 & 58.3$\pm$11.3 \\
        \hline
        Social Security & 72 & 64.2$\pm$11.1 & \underline{\textbf{69.3$\pm$10.7}} & 65.0$\pm$11.0 & 57.8$\pm$11.4 & 54.5$\pm$11.5 \\
        \hline
        Fathers & 45 & \textbf{38.5$\pm$14.2} & 37.8$\pm$14.2 & 37.8$\pm$14.2 & 13.3$\pm$9.9 & 26.9$\pm$13.0 \\
        \hline
        Family & 30 & \underline{\textbf{72.7$\pm$15.9}} & 55.6$\pm$17.8 & 48.0$\pm$17.9 & 60.0$\pm$17.5 & 33.3$\pm$16.9 \\
        \hline
        Superfund & 19 & \textbf{84.6$\pm$16.2} & 73.3$\pm$19.9 & 73.3$\pm$19.9 & 45.5$\pm$22.4 & 73.3$\pm$19.9 \\
        \hline
    \end{tabular}
    \caption{$F_1$ for for testing on one topic when training on all other paragraphs; D1 vs. \{D0, T0\}.}
    \label{tab:file-topics-noneasy}
  \centering
    \vspace{-6mm}%Put here to reduce too much white space after your table 
\end{table}

\subsection{Efficacy of Keyword Searching vs. Machine Learning}
Tables~\ref{tab:custodian-attorney-category-a} through~\ref{tab:file-topics-noneasy} consistently indicate that classifiers based on matching any one of a manually selected set of keywords were consistently outperformed by classifiers that learned from training examples. This result is consistent with numerous other studies~\cite{grossman2010technology,baron2006trec}, suggesting either that human designers lack the sufficient insight into vocabulary used in the items that are sought in the collection, or perhaps that such ``flat'' disjunctive classification rules are not sufficiently expressive. It may, however, ultimately be possible to augment human performance at this task by suggesting terms for human designers to consider.  As an example, Table~\ref{tab:top-words} shows the keywords with the greatest positive and negative weights for the exempt class for one fold of the first logistic regression classifier shown in Table~\ref{tab:custodian-attorney-category-a}. Strikingly, there is only one word (option) in common between the keywords chosen by Reviewer A before annotating anything and the words on which a classifier most heavily relies after training on Reviewer A's annotations.

\begin{table}[h]
  \centering
    \begin{tabular}{|c|c|}
        \hline
        \textbf{Top words, positive weights} & \textbf{Top words, negative weights}\\
        \hline
        option & today\\
        counter & committee\\
        options & state \\
        doj & subject \\
        authority & clinton \\
        program & education \\
        increase & human \\
        splitting & family \\
        idea & let \\
        largest & police \\
        accreditation & 30 \\
        language & soon \\
        initiatives & eop \\
        coordinator & radiation \\
        vouchers & 00 \\
        necessary & experiments \\
        targeted & prisoner \\
        significant & approved \\
        think & 18 \\
        action & jose \\
        \hline
    \end{tabular}
    \caption{Top 20 words with the greatest positive and negative weights for LR in Table~\ref{tab:custodian-attorney-category-a}, leftmost case, first fold.}
    \label{tab:top-words}
  \centering
    \vspace{-6mm}%Put here to reduce too much white space after your table 
\end{table}

\section{Discussion}

These experiments indicate that using supervised machine learning to help human reviewers identify content exempt from release under the FOIA Exemption 5 deliberative process privilege is feasible.  As it stands, human reviewers manually review records without the records being “weighted” or ranked in any fashion. Applying a classifier would introduce at least three efficiencies into the existing process.  First, it could be used to push to the front of a large-scale review process those records least likely to contain exempt material, allowing reviewers to focus first on the portion of the collection that could perhaps be reviewed most rapidly. Second, for any portions of records that the classifier determines contain potentially withholdable material, highlighting could help to draw the reviewer's eye to passages that would benefit from the most careful review. Third, once the review has been completed, a classifier could be trained from all of the resulting decisions and then used to detect inconsistencies in those decisions~\cite{ishita2020using}.  Such a machine-assisted final review process could help to further improve confidence in the review process, possibly catching both inadvertent errors of commission (incorrect release) and errors of omission (incorrect withholding). Sampling and independent review can then be used as a final step to still further increase confidence in the process that was used in the review.

The results in Tables~\ref{tab:custodian-attorney-category-a} through~\ref{tab:custodian-attorney-category-d} clearly indicate that classifiers can be trained to reliably highlight the vast majority of content that is exempt from release, and that they can do so with sufficient precision to be directly useful.  As with many applications of supervised machine learning, closely matching the training and test conditions, as in condition A, yields the best results.  This suggests that iterative retraining of the classifier as more annotations are made would be a promising option to explore.  

As the results in Tables~\ref{tab:easy-zeros} and in~\ref{tab:non-trivial}, further improvements might be obtained by incorporating specialized classifiers for entire documents that are categorically non-exempt (what we have called ``easy zeros'') or for document elements that human reviewers would not need to spend time reviewing (what we have called ``trivial zeros'').  These additional classifiers could make use of features from, for example, layout analysis, genre detection, or authorship attribution.  Future work should also explore whether those additional features and others (e.g., from opinion detection) might help to improve our base classifiers' abilities to make the most challenging decisions (between D1 and D0).

%\jason {THESE WERE THE OBSERVATIONS I COULDNT FIND ON OUR CALL FOR ALL OF US TO FLESH OUT FURTHER:
%Additional observations from Section 4 Experimental Results: telling the best story in terms of the results; what achieving modest improvements over the baseline of All 1's really translates into in the real world; thoughts about how to improve performance of the classifier; the meaning of "E0"in Batch 4 and some comment about ideally we would wish to build a classifier that is able to distinguish as between inter- and intra-agency documents and those that fall outside the threshold test for b5}

Our findings are subject to a number of caveats.  First, this research was conducted on a small scale involving the records primarily of one senior official in government, coupled with a limited number of records from a second official.  Before a classifier could be considered reliable, it would need to be shown that based on a given training set of records, the classifier could perform well across a range of records drawn from a greater set of key custodians in federal agencies thought to be holding responsive records to a given FOIA request.

Second, our findings are based on imposing a measure of artificiality in that paragraphs were manually identified for purposes of facilitating comparisons between human and machine annotations.  Our experiments with Begin-Inside-Outside classification point the way toward more flexible approaches that could work at the scale of sentences, lines, or even words.

Third, although we have adopted the opinions of experienced reviewers as the gold standard for our experiments, there will always be room for human judgment in the process and necessarily a measure of uncertainty in the results.  The key question to ask, therefore, is not how good is our classifier, as if it were operating on its own, but rather how useful is our classifier when used to support a real FOIA review process, by real FOIA reviewers who must grapple with the discretion and degree of ambiguity embodied in FOIA law.  The prospect of automation should always be considered to be part of what is ultimately going to remain a human-centered review process.  A natural next step would therefore be to study the use of systems that include classifiers like those that we have described here on a larger universe of records. McDonald et al. studied the effect of the accuracy of sensitivity classification on human reviewers performance~\cite{mcdonald2020accuracy}. Results show that quality of classification results affects reviewer's performance in terms of correctness and time to finish. We would like to perform a similar study and work closely with FOIA reviewers to understand what the best way is to couple their efforts with sensitivity classification.

Finally, we note that our experiments here have focused on the deliberative process privilege, which is one part of one exemption in the federal FOIA law.  Our promising results suggest that further work on other exemptions at the federal and state levels, and on other aspects of FOIA (such as judging the relevance of a document to a request) justifiably deserve attention.   

\section{Conclusion and Future Work}

This study has shown that classifiers trained using supervised machine learning can potentially be of benefit in highlighting portions of records that are within the scope of the deliberative process privilege under FOIA Exemption 5.  Because considerable time and resources are currently devoted to manually reviewing records in responding to FOIA requests, including with respect to the statutory duty to reasonably segregate exempt and non-exempt material, automated ways with which to identify potentially exempt portions of text would be welcome.  The use of such methods would not obviate the need for human review; rather, efficient and accurate ``flagging'' of material deemed by a classifier to be potentially within scope of the privilege would assist reviewers in determining which records to review on the ``front end'' of any overall review effort.

Although our focus has been on training classifiers to detect deliberative process privilege material, this research exercise provides a path forward for applying a similar set of classifiers, or possibly an ensemble of them, to other FOIA exemptions.  For example, FOIA Exemption 4 allows for withholding records containing trade secrets and confidential or proprietary information.  FOIA Exemption 5 also allows for the withholding of attorney-client privileged information, as well as attorney work product.  FOIA Exemption 6 allows records containing forms of PII to be withheld.  FOIA Exemption 7 allows for withholding of various categories of law enforcement records the release of which would constitute an unwarranted invasion of privacy.   All of these exemptions (and possibly others) could potentially benefit from automated classifiers being used to identify records or portions of records that might be deemed withholdable.  

The need for automated classification for purposes of satisfying FOIA obligations is growing, especially  given the digital turn that the records of our government are currently undergoing.   After 2022, NARA will require agencies to transfer permanent records to the archives only in electronic form.  Additionally, after 2022 agencies will be expected to manage all of their records (temporary and permanent) in electronic form~\cite{transition2019}.   Since 2016, e-mail records have been required to be similarly managed.  With the introduction of what is known as the ``Capstone'' policy for e-mail retention~\cite{archives2015capstone}, tens or hundreds of millions of e-mail records, including their many attachments, will soon be residing in any number of the larger agencies' repositories, all potentially subject to FOIA requests.   Absent automated ways to reliably  filter exempt from non-exempt materials to comply with FOIA obligations, access to a growing percentage of the government's records will be diminished.   Any method that promises to facilitate the needed document review should be worthy of serious consideration.

Subject to the above caveats, the prospect of having reliable machine learning methods in place for assisting human reviewers in making FOIA determinations means that the documentary heritage embodied in government records will be made more accessible to the American people.  This will continue to be of great importance in an ever-expanding universe of public records, both during and after the COVID-19 era.  

\begin{acks}
We wish to thank Patricia Weth, a senior government attorney with substantial expertise in FOIA law, for her assistance in annotating documents.   We also wish to thank Dana Simmons, a supervisory archivist at the Clinton Library, for her assistance in helping identify a suitable test collection within the Library’s holdings.  This work was supported in part by NSF grant IIS-1618695.
\end{acks}

\bibliographystyle{ACM-Reference-Format}
\bibliography{sample-base}

\end{document}